\documentclass[10pt,conference,letterpaper]{IEEEtran}
\newcommand{\TWOCOLUMN}[1]{#1}\newcommand{\ONECOLUMN}[1]{}
\IEEEoverridecommandlockouts

\newcommand{\paperlink}[2]{}
\newcommand{\confpaperlink}[2]{}
\usepackage{setspace}
\usepackage{cite}
\usepackage[pdftex]{graphicx}
\usepackage[cmex10]{amsmath}
\usepackage{amssymb}

\usepackage{mathtools}
\usepackage{texdef2015}
\usepackage{url}
\usepackage{epstopdf}
\usepackage{fixltx2e}
\usepackage{tikz}
\usepackage{tikz}
\usetikzlibrary{automata,arrows,positioning,calc,%
shapes,chains}

\newtheorem{theorem}{Theorem}

\newcommand{\nstack}[2]{\begin{matrix} {\mathbf #1}\\ (#2)\end{matrix}}

\newcommand{\Lcal}{\mathcal{L}}
\newcommand{\Qcal}{\mathcal{Q}}

\newcommand{\R}{\mathbb{R}}
\newcommand{\dq}[2][q]{\delta_{#2,#1}}
\newcommand{\dqt}[2][q(t)]{\delta_{#2,#1}}
\newcommand{\ql}{q_l}

\newcommand{\laml}[1][l]{\lambda^{(#1)}}

\newcommand{\cvec}[1]{\Bracket{#1}}

\newcommand{\smvec}[1]{\left[\begin{smallmatrix} #1\end{smallmatrix}\right]}
\newcommand{\rowvec}[1]{[\begin{matrix} #1\end{matrix}]}
\newcommand{\rvec}[1]{\smrvec{#1}}
\newcommand{\smrvec}[1]{\setlength\arraycolsep{2pt}\rowvec{#1}}
\newcommand{\avgage}{\tilde{\age}}

\newcommand{\piv}{\text{\boldmath{$\pi$}}}
\newcommand{\pivbar}{\text{\boldmath{$\bar{\pi}$}}}
\newcommand{\pibar}{\bar{\pi}}
\newcommand{\vvbar}{\bar{\vv}}

\renewcommand{\vec}[1]{\begin{bmatrix} #1\end{bmatrix}}
\renewcommand{\bv}{\mathbf{b}}

\newcommand{\age}{\Delta}

\newlength{\swwidth}
\newcommand{\stw}[2]{\text{\settowidth{\swwidth}{\text{\ensuremath{#1}}}\makebox[\swwidth]{#2}}}
\newcommand{\vz}[1][0]{\stw{v_{00}}{#1}}
\newcommand{\xz}[1][0]{\stw{x_{0}}{#1}}

\newcommand{\DotM}{$\cdot$/M}

\title{Age of Information in a Network of Preemptive Servers}
\author{\IEEEauthorblockN{Roy D.~Yates
\thanks{This work was supported by NSF award CNS-1422988 and will be presented at the 2018 IEEE Infocom  Age of Information workshop.}
}
\IEEEauthorblockA{WINLAB, Department of Electrical and Computer Engineering\\
Rutgers University\\
{\em ryates@winlab.rutgers.edu}}
}

\begin{document}
\maketitle
\begin{abstract}
A source submits status updates to a network for delivery to a destination monitor. Updates follow a route through a series of network nodes. Each node is a last-come-first-served queue supporting preemption in service. 
We characterize the average age of information at the input and output of each node in the route induced by the updates passing through. For Poisson arrivals to a line network of preemptive memoryless servers, we show that average age accumulates through successive network nodes. 
 \end{abstract} 
\section{Introduction}
The need for timely knowledge of the system state in remote monitoring applications has led to the development and analysis of status update age metrics. 
When randomly arriving updates are queued in a service facility, early work \cite{KaulYatesGruteser-Infocom2012,%
KamKompellaEphremides2014ISIT} showed that it is generally in the self-interest of a source to limit its offered load. In particular,  the updating must balance between too infrequent updates and overly frequent updates that induce  queueing delays. 

This observation has prompted the study of age in lossy systems that discard updates to avoid building queues. These include the last-come-first served (LCFS) queue with preemption either in waiting or service \cite{KaulYatesGruteser-Infocom2012,2012CISS-KaulYatesGruteser,2012ISIT-YatesKaul} and packet management mechanisms that restrict the the number of queued packets \cite{CostaCodreanuEphremides2014ISIT} or discard waiting packets as they become stale \cite{KamKompellaNWE-ISIT2016}. However, these contributions consider only single hop communication systems. 

Recently, there have been effort to examine age in multihop network settings \cite{BedewySunShroff-ISIT2017,BedewySunShroff17,%
TalakKaramanModiano-Allerton2017}.
In particular, this work is closely related to the Last-Generated-First-Served (LGFS) multihop networks studied in \cite{BedewySunShroff-ISIT2017,BedewySunShroff17}. When update transmission times over network links are exponentially distributed,  sample path arguments were used to show that a preemptive Last-Generated, First-Served (LGFS) policy results in smaller age processes at all nodes of the network than any other causal policy. However, these structural results do not facilitate the explicit calculation of age. 

In this work, we consider preemptive LCFS servers, a special case of the LGFS discipline, in the multihop line network shown in Figure~\ref{fig:network}.  For Poisson arrivals and memoryless preemptive servers, we use a stochastic hybrid systems (SHS) model of the age processes in the network to derive a simple expression for the average age at each node. 

This work was motivated by a simple question regarding the line network with $n=2$ nodes and service rates $\mu_1$ and $\mu_2$. The network traffic depends qualitatively on these service rates $\mu_1$ and $\mu_2$. For example if $\mu_1\ll\lambda\le \mu_2$, packet dropping will occur primarily at node $1$ but traffic will pass quickly, and with relatively little dropping, through node $2$, On the other hand, with  $\mu_1\gg\lambda\ge\mu_2$ there will be substantial dropping at node $2$, but little dropping at node $1$.   From the perspective  of average age at the monitor, is it better for dropping to occur earlier ($\mu_1<\mu_2$) or later ($\mu_1>\mu_2$) in the network?  A reasonable hypothesis is that $\mu_1< \mu_2$ wastes the resources of 
the faster downstream server.  In fact, we will see in \Thmref{line-age} that average age at the monitor is insensitive to the ordering of the servers.

In this work, the network model is described in Section~\ref{sec:model}. This is followed in Section~\ref{sec:SHS} with a summary of results on  stochastic hybrid systems (SHS)  for age analysis from \cite{YatesKaul2017arxiv} that will be the basis for our age analysis. In Section~\ref{sec:tandem}, we apply SHS to the $2$-node tandem queue. We  use a $4$-state model that describes the occupancy of each server in the network. While this analysis may be instructive, it is shown in Section~\ref{sec:fake}  that the preemptive service facilitates an analysis using a ``fake updates'' technique from \cite{YatesKaul2017arxiv}  that reduces the discrete state space to just one state. Some simulation experiments are provided in Section~\ref{sec:plots} and conclusions appear in Section~\ref{sec:conclusion}.

\section{System Model}\label{sec:model}
We model the updating process as a source that submits update packets as a rate $\lambda$ Poisson process to a network.  As depicted in Figure~\ref{fig:network}, updates follow a route through $n$ nodes in a network to a monitor. At node $i$, an update has an exponential $(\mu_i)$ service time, independent of the arrival process and  service times at other nodes.  However, we forgo queueing in this network; each node is a \DotM/1/1 preemptive server. Upon arrival at a server, an update immediately goes into service and any  update currently in service  is preempted and discarded.

This is a useful model when the time an update spends in the head-of-line position in the network interface is dominated by waiting for transmission. For example, in a congested wireless network, the service time would be dominated by the  MAC access delay and the transmission time of the packet is negligible. 
In this case, it would be feasible and desirable for the head-of-line update packet, i.e. the nominal update in service,  to be preempted by a more recent arrival. 

Starting at time $t=0$, the source submits  status updates at successive times $U_1,U_2\ldots$ such that update $i$ submitted at time $U_i$ is delivered at time $U_i+T_i$. The $T_i$ are dependent random variables. Moreover, because of preemption,  the line network is lossy and $T_i=\infty$ for those updates that are discarded.
At time $t$, the most recent received update is time-stamped
$U(t)=\max\set{U_i| U_i+T_i\le t}$ 
and thus 
the \emph{status update age}, which we  refer to as simply the \emph{age}, is $\age(t)=t-U(t)$.
The system performance is given by the average {\em status update age} $\age=\limty{t}\E{\age(t)}$.


For the $n$-node network in Figure~\ref{fig:network}, we denote the average age at the monitor by $\age(\lambda,\mu_1,\cdots,\mu_n)$. In the case of $n=1$ node, we have a simple M/M/1/1 preemptive queue. This has also been called the Last-Come First-Served with preemption in service (LCFS-S) queue \cite{YatesKaul2017arxiv}. Using a graphical approach, it was shown \cite{2012CISS-KaulYatesGruteser} that the time average of $\age(t)$ approaches
\begin{equation}\eqnlabel{age1node}
\age(\lambda,\mu_1)=\frac{1}{\lambda}+\frac{1}{\mu_1}.
\end{equation}
In this prior work, an end-to-end network is modeled as a single service facility. The key analytical steps involve the the system time $T$ and the interarrival time $Y$ of delivered packets and the challenge is the computation of the correlation $\E{TY}$. However, when the network has $n\ge 2$ nodes, the graphical method fails because the queueing of updates in the network is non-trivial for several reasons:
\begin{figure}[t]
\centering
\begin{tikzpicture}[->, >=stealth', auto, semithick, node distance=1.5cm]
\tikzstyle{every state}=[fill=white,minimum size=28pt,
draw=black,thick,text=black,scale=1]
\node[state]    (0)                     {\shortstack{\small Source\\ $\lambda$}};
\node[state]    (1)[right of=0]   {$\mu_1$};
\node[state]    (2)[right of=1]   {$\mu_2$};
\node[state,draw=white] (H1)[right of=2] {};
\node[state] (n)[right of=H1]{$\mu_{n}$};
\node[state,scale=1] (monitor)[right of=n] {\small Monitor};
\path
(0) 	edge (1) 
(1) edge  (2)
(2)     edge  (H1)  
(H1) edge (n) 
(n) edge (monitor);
\end{tikzpicture}
\caption{The $n$-node  line network model. Node $i$ is a rate $\mu_i$ \DotM/1/1 preemptive server.}
\label{fig:network}
\end{figure}
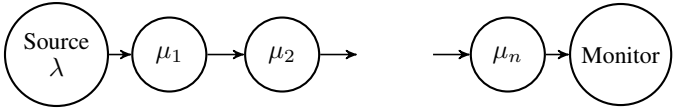
\begin{itemize}
\item The line network is lossy as updates are discarded when they are preempted. Even the calculation of the effective arrival rate of updates at the monitor is challenging.  
\item The departure process at each node, even node $1$,  is not memoryless.  
\item The arrivals at node $2$ and subsequent nodes are not fresh; instead they are aged by their passage  through prior nodes. The age of a packet arriving at node $i$ may be correlated with its service times at preceding nodes. That is, the interarrival time of an update  may be correlated with its age.
\item Updates that reach the monitor may be subject to a survivor bias as each was lucky enough in its service times to avoid being preempted. 
\end{itemize}   

\section{Stochastic Hybrid Systems for AoI}\label{sec:SHS}
Because of the complexity of the lossy queueing process in the line network, we take a different non-graphical approach to average age analysis. Following \cite{YatesKaul2017arxiv}, we model the system as a stochastic hybrid system (SHS) with hybrid state $(q(t),\xv(t)$). In the SHS model,
$q(t)\in\Qcal$ is  discrete and typically describes the Markov state of the queueing system while the row vector $\xv(t)\in\R^{n+1}$ is continuous and captures the evolution of a collection of age-related processes. 

In general, SHS is a powerful modeling framework with many variations \cite{hespanha2006modelling,teel2014stability}. In this work, we use a simplified form of SHS for AoI analysis introduced in \cite{YatesKaul2017arxiv} in which $\xv(t)$ is a piecewise linear process. In the interest of completeness, we now summarize the basics of this simplified SHS; further details can be found in \cite{YatesKaul2017arxiv} and references therein. 

In the graphical representation of the Markov chain $q(t)$, each state $q\in\Qcal$ is a node and each transition $l\in\Lcal$ is a directed edge $(q_l,q'_l)$ with transition rate $\laml\dqt{q_l}$. Note that the Kronecker delta function $\dq{q_l}$ ensures that transition $l$ occurs only in state $q_l$. For each state $q$, we define 
\begin{align}\eqnlabel{Lcalqbar}
\Lcal'_{q}&=\set{l\in\Lcal: q'_l=q}, &
\Lcal_{q}&=\set{l\in\Lcal: q_l=q}
\end{align}
as the respective sets of incoming and outgoing  transitions. For each transition $l$, there is  transition reset mapping that can induce a discontinuous jumps in the continuous state $\xv(t)$. For AoI analysis, we employ a linear mapping 
of the form $\xv'=\xv\Amat_l$.  That is, transition $l$ causes the system to jump to discrete state $q'_l$ and resets the continuous state from $\xv$ to $\xv'=\xv\Amat_l$.   Moreover,  in each discrete state $q(t) = q$, the continuous state evolves according to 
$\dot{\xv}(t)=\bv_q$.

In using a piecewise linear SHS for AoI, the elements of $\bv_q$ will be binary. We will see that the ones in $\bv_q$ correspond to certain relevant components of $\xv(t)$ that grow at unit rate in state $q$ while the zeros mark components of $\xv(t)$ that are irrelevant in state $q$ to the age process and need not be tracked.
For tracking of the age process, the transition reset maps are binary: $\Amat_l\in\set{0,1}^{(n+1)\times (n+1)}$. The linear mappings $\Amat_l$ will depend on the specific network system and the indexing scheme for updates in the system.

The transition rates $\laml$ correspond to the transition rates associated with the continuous-time Markov chain for the discrete state $q(t)$; but there are some differences. Unlike an ordinary continuous-time Markov chain, the SHS may include self-transitions in which the discrete state is unchanged because a reset occurs in the continuous state. Furthermore, for a given pair of states $i,j\in\Qcal$, there may be multiple transitions $l$ and $l'$ in which the discrete state jumps from $i$ to $j$ but the transition maps $\Amat_l$ and $\Amat_{l'}$ are different.

It will be sufficient for average age analysis 
to define for all $\qhat\in\Qcal=\set{0,1,\ldots,m}$,
\begin{subequations}\eqnlabel{pivi-defns}
\begin{align}
\pi_{\qhat}(t) &= 
\E{\dqt{\qhat}},
\eqnlabel{piqhat-defn}\\
v_{\qhat j}(t)&=
\E{x_j(t)\dqt{\qhat}},\quad  j\in\set{0,\ldots,n},
\eqnlabel{vqi-defn}\\
\shortintertext{and the vector functions}
 \vv_{\qhat}(t)&=\cvec{v_{\qhat 0}(t),\dots,v_{\qhat n}(t)}=\E{\xv(t)\dqt{\qhat}}. \eqnlabel{vv-defn}
 \end{align}
\end{subequations}

We note that $\pi_{\qhat}(t)$ denotes the discrete Markov state probabilities, i.e.,
\begin{equation}
\pi_{\qhat}(t)=\E{\dqt{\qhat}}=\prob{q(t)=\qhat}.
\end{equation} 
Similarly, $\vv_{\qhat}(t)$ measures correlation between the age process $\xv(t)$ and the occupancy of the discrete state $q(t)$.

A fundamental assumption for age analysis is that the Markov chain $q(t)$ is ergodic; otherwise, time-average age analysis makes little sense. Under this assumption, the state probability vector $\piv(t)=\rvec{\pi_0(t) &\cdots&\pi_m(t)}$ always  converges to the unique stationary vector $\bar{\piv}=\rvec{\pibar_0&\cdots&\pibar_m}$ satisfying
\begin{subequations}
\begin{align}
\bar{\pi}_{\qbar}\sum_{l\in\Lcal_{\qbar}}\laml&=\sum_{l\in\Lcal'_{\qbar}}\laml\bar{\pi}_{\ql},\quad \qbar\in\Qcal,
\eqnlabel{AOI-SHS-pi}\\
\sum_{\qbar\in\Qcal}\bar{\pi}_\qbar&=1.
\end{align}
\end{subequations}
When $\piv(t)=\bar{\piv}$,  it is shown in \cite{YatesKaul2017arxiv} that $\vv(t)=\rvec{\vv_0(t)&\cdots&\vv_m(t)}$ obeys the system of first order differential equations such that for all $\qbar\in\Qcal$,
\begin{align}
\dot{\vv}_{\qbar}(t)&=
\bv_{\qbar}\pibar_{\qbar}+\sum_{l\in\Lcal'_{\qbar}}\laml \vv_{\ql}(t)\Amat_l
-\vv_{\qbar}(t)\sum_{l\in\Lcal_{\qbar}}\laml.
\eqnlabel{vv-derivs-pibar}
\end{align}
Depending on the  reset maps $\Amat_l$, the differential equation \eqnref{vv-derivs-pibar} may or may not be stable. However,  when \eqnref{vv-derivs-pibar} is stable, $\dot{\vv}(t)\to0$ and 
each $\vv_{\qbar}(t)=\E{\xv(t)\dqt{\qbar}}$ converges to a limit $\vvbar_{\qbar}$ as $t\goes\infty$. 
In this case, it  follows 
that
\begin{align}
\E{\xv}&\equiv\limty{t}\E{\xv(t)}
\TWOCOLUMN{\nn &}=\limty{t}\sum_{\qbar\in\Qcal} \E{\xv(t)\delta_{\qbar,q(t)}}
=\sum_{\qbar\in\Qcal} \bar{\vv}_{\qbar}.
\end{align}
Following the convention in \cite{YatesKaul2017arxiv} that $x_0(t)=\age(t)$ is the age at the monitor, the average age of the process of interest is then
$\age=\E{x_0}=\sum_{\qbar\in\Qcal}\vbar_{\qbar 0}$.
The following theorem provides a simple way to calculate the average age in an ergodic queueing system.
\begin{theorem}\thmlabel{AOI-SHS}
\cite[Theorem~4]{YatesKaul2017arxiv}
If the discrete-state Markov chain $q(t)$ is ergodic with stationary distribution $\bar{\piv}$ and we can find a non-negative solution $\vvbar=\rvec{\vvbar_0&\cdots\vvbar_m}$ 
such that 
\begin{subequations}
\begin{align}
\bar{\vv}_{\qbar}\sum_{l\in\Lcal_{\qbar}}\laml &=\bv_{\qbar}\bar{\pi}_{\qbar}+ \sum_{l\in\Lcal'_{\qbar}}\laml \bar{\vv}_{\ql}\Amat_l,\qquad \qbar\in\Qcal,\eqnlabel{AOI-SHS-v}
\end{align}
then
the differential equation \eqnref{vv-derivs-pibar} is stable and 
the average age of the AoI SHS is given by
\begin{equation}
\age=\sum_{\qbar\in\Qcal} \vbar_{\qbar 0}.
\eqnlabel{age-vsum}
\end{equation}
\end{subequations}
\end{theorem}
\section{Age in the Tandem Queue with Preemption}\label{sec:tandem}
We now use \Thmref{AOI-SHS} to evaluate the age $\age(\lambda,\mu_1,\mu_2)$ at the monitor in the line network of Figure~\ref{fig:network} with $n=2$ intermediate nodes. This example will demonstrate how to use \Thmref{AOI-SHS} in a straightforward way to evaluate average age in a network. 

 In the $2$-node network, the set of discrete states is $\Qcal=\set{00,10,01,11}$ such that for $q_1q_2\in\Qcal$, $q_i=1$ indicates that node $i$ is serving an update packet. It will also be convenient to refer to the states  $\Qcal=\set{0,1,2,3}$ such that $q=q_1+2q_2\in\Qcal$. 

The continuous state is $\xv(t)=\rvec{x_0(t),x_1(t),x_2(t)}$ such that $x_0(t)$ is the age at the monitor, and, when there is an update in service at node $i$, $x_i(t)$ is the age of that update. When node $i$ is idle, $x_i(t)$ is irrelevant and we set $x_i(t)=0$. In particular, in any state $q$ in which node $i$ is idle, we hold $x_i(t)=0$  while in that state. Otherwise, if node $i$ is serving an update in a  state $q$, then $x_i(t)$ increases at unit rate in that state.  It follows that in state $q$ we set
\begin{equation}\eqnlabel{b-2nodes}
\bv_q=\begin{cases}
\rvec{1&0&0} & q=0,\\
\rvec{1&1&0} & q=1,\\
\rvec{1&0&1} & q=2,\\
\rvec{1&1&1} & q=3.
\end{cases}
\end{equation}
%
It follows from \eqnref{b-2nodes} that in each $q$, the age  at the monitor, $x_0(t)$, grows at unit rate. On the other hand, in states $q\in\set{1,3}$, node $1$ is serving an update whose age $x_1(t)$ is growing at unit rate.  Similarly,  in states $q\in{2,3}$, node $2$ is serving an update whose age $x_2(t)$ is growing at unit rate. 
\begin{figure}[t]
\centering
\begin{tikzpicture}[->, >=stealth', auto, semithick, node distance=2.25cm]
\tikzstyle{every state}=[fill=white,draw=black,thick,text=black,scale=1]
\node[state]    (0)                     {$\nstack{0}{00}$};
\node[state]    (2)[right of=0]   {$\nstack{2}{01}$};
\node[state]    (1)[above of=2]   {$\nstack{1}{10}$};
\node[state] (3)[right of =2] {$\nstack{3}{11}$};
\path
(0) 	edge[bend left=20,above]     node{$1$}  (1)
(1)  edge[loop right,right]  node {$2$} (1)
(1) edge[bend left=20, right] node {$3$} (2)
(2) edge[bend left=20,below] node {$4$} (0)
(2) edge[bend left=20,above] node {$5$} (3)
(3) edge[bend right=20,above] node {$6$} (1)
(3) edge[bend left=20,below] node {$7$} (2)
(3) edge[loop right,right] node {$8$} (3);
\end{tikzpicture}
\caption{The SHS Markov chain for the line network with $n=2$ nodes.
The transition rates and transition/reset maps for links $l=1,\ldots,8$ are shown in Table~\ref{tab:MC-2nodes}.}
\label{fig:MC-2nodes}
\end{figure}
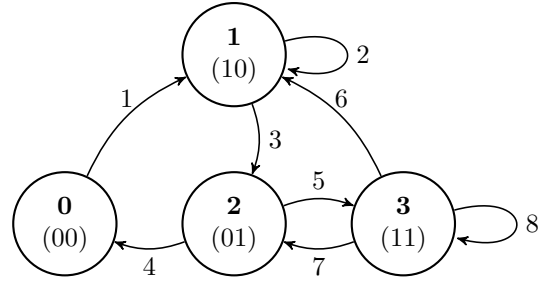

The Markov chain for the occupancy of the network is shown in Figure~\ref{fig:MC-2nodes}. The edges are labeled $l\in\set{1,2,\ldots,8}$. For each transition $l$, Table~\ref{tab:MC-2nodes} lists   the state transition pair $q_l\to q'_l$, the transition rate $\laml$, the transition mapping $\xv'=\xv\Amat_l$,  the matrix $\Amat_l$ and, to facilitate using \Thmref{AOI-SHS}, $\vv_{q_l}\Amat_l$. 

\begin{table}[t]
\begin{displaymath}
\begin{array}{rrcccc}
l & q_l\to q'_l & \laml & \xv\Amat_l &\Amat_l&\vvbar_{q_l}\Amat_l\\\hline
1 & 0\to 1 &\lambda 	& \rvec{x_0& \xz&\xz}&\smvec{1 & 0&0\\ 0 & 0&0\\0&0&0} &\rvec{\vbar_{00}&\vz&\vz}\\[0.5em]
2 & 1\to 1 & \lambda	& \rvec{x_0&\xz&\xz}&\smvec{1&0&0\\ 0&0&0\\ 0&0&0}
&\rvec{\vbar_{10}&\vz&\vz}\\[0.5em]
3 & 1\to2 & \mu_1		& \rvec{x_0&\xz&x_1}&\smvec{1&0&0\\ 0&0&1\\ 0&0&0}
&\rvec{\vbar_{10}&\vz&\vbar_{11}}\\[0.5em]
4 & 2\to0& \mu_2 	& \rvec{x_2&\xz&\xz}&\smvec{0&0&0\\ 0&0&0\\ 1&0&0}
& \rvec{\vbar_{22}&\vz&\vz}\\[0.5em]
5 & 2\to3& \lambda& \rvec{x_0&\xz&x_2}&\smvec{1&0&0\\ 0&0&0\\ 0&0&1}
&\rvec{\vbar_{20}&\vz&\vbar_{22}}\\[0.5em]
6 & 3\to1&  \mu_2 & \rvec{x_2&x_1&\xz}&\smvec{0&0&0\\ 0&1&0\\ 1&0&0}
&\rvec{\vbar_{32}&\vbar_{31}&\vz}\\[0.5em]
7 & 3\to2& \mu_1	&\rvec{x_0&\xz&x_1} &\smvec{1&0&0\\ 0&0&1\\ 0&0&0}
&\rvec{\vbar_{30} &\vz &\vbar_{31}}\\[0.5em]
8 & 3\to3& \lambda & \rvec{x_0&\xz&x_2} &\smvec{1&0&0\\ 0&0&0\\ 0&0&1} &\rvec{\vbar_{30}&\vz&\vbar_{32}}
\end{array}
\end{displaymath}
\caption{Table of transitions for the Markov chain in Figure~\ref{fig:MC-2nodes}.}\label{tab:MC-2nodes}
\vspace{-5mm}
\end{table}
We now describe the transitions $l$. We note that the age $x_0(t)$ at the monitor changes only in those transitions in which an update completes service at node $2$ and is delivered to the monitor. Corresponding to Table~\ref{tab:MC-2nodes}, the transitions are:
\begin{enumerate}
\item 
In an idle network, a fresh update arrives at node $1$. The age $x'_0=x_0$ at the monitor is unchanged, $x'_1=0$ because the arrival is fresh, and $x'_2=0$ because $x_2$ is irrelevant in state $1$. 
\item 
In state $1$, a fresh update arrives and preempts the update in service at node $1$.  The age $x'_0=x_0$ at the monitor is unchanged, $x'_1=0$ because the arrival is fresh, and $x'_2=0$ because $x_2$ is irrelevant in state $1$. 
\item 
In state $1$, the update at node $1$ completes service and moves to node $2$. The age $x'_0=x_0$ at the monitor is unchanged, $x'_1=0$ because $x_1$ becomes irrelevant in state $2$ and $x'_2=x_1$ because the update now at node $2$ is the update that was previously at node $1$.
\item 
In state $2$, the update at node $2$ completes service and is delivered to the monitor. The system moves to state $0$. The age at the monitor becomes $x'_0=x_2$. In addition, $x'_1=x'_2=0$ since $x_1$ and $x_2$ are irrelevant in state $0$. 
\item 
The arrival of a fresh update  at node $1$ induces the $2\to 3$ state transition. The age $x'_0=x_0$ is unchanged. At node $1$, $x'_1=0$ because the update is fresh. At node $2$, the age $x'_2=x_2$ is unchanged. 

\item 
The update at node $2$ completes service and is delivered to the monitor. The system moves to state $1$. The age at the monitor becomes $x'_0=x_2$. In addition, $x'_1=x_1$ is unchanged because the update at node $1$ remains in place and  $x'_2=0$ since $x_2$ are irrelevant in state $1$. 
\item 
The update at node $1$ completes service and moves to node $2$, preempting the update that had been in service at node $2$. The age $x'_0=x_0$ is unchanged, $x'_1=0$ because $x_1$ becomes irrelevant in state $2$, and $x'_2=x_1$ because the update now at node $2$ is the update that was previously at node $1$.
\item 
A fresh update arrives at node $1$, preempting the update that had been in service at node $1$. The age $x'_0=x_0$ is unchanged, $x'_1=0$ because the new update at node $1$ is fresh, and $x'_2=x_2$ is unchanged because the update at node $2$ stays in place.
\end{enumerate}
To use \Thmref{AOI-SHS} to find the average age, we first find the stationary probabilities $\pivbar$. It is straightforward to verify that
\begin{subequations}\eqnlabel{pibar-2nodes}
\begin{align}
\pibar_0&= \frac{\mu_1\mu_2}{(\mu_1+\lambda)(\mu_2+\lambda)},&
\pibar_2&=\frac{\lambda}{\mu_2}\pibar_0,\\
\pibar_1&=\frac{\lambda}{\mu_1}\paren{\frac{\mu_1+\mu_2+\lambda}{\mu_1+\mu_2}}\pibar_0,
 &
\pibar_3&=\frac{\lambda^2}{\mu_2(\mu_1+\mu_2)}\pibar_0.
\end{align}
\end{subequations}

Second, we need to find a non-negative $\vvbar$ satisfying \eqnref{AOI-SHS-v}. Defining $\alpha_i=\lambda+\mu_i$ for $i=1,2$, and $\alpha_{3}=\lambda+\mu_1+\mu_2$, it follows from \eqnref{AOI-SHS-v} and Table~\ref{tab:MC-2nodes} that
\begin{subequations}\eqnlabel{AOI-SHS-v-2nodes}
\begin{align}
\vvbar_0\lambda &=\pibar_0\bv_0+\mu_2\rvec{\vbar_{22}&0&0},
\eqnlabel{v-2nodes-q0}\\
\vvbar_1\alpha_1&=\pibar_1\bv_1
+\lambda\rvec{\vbar_{00}&0&0}
\nn&\qquad\qquad
+\lambda\rvec{\vbar_{10}&0&0}
+\mu_2\rvec{\vbar_{32}&\vbar_{31}&0},\eqnlabel{v-2nodes-q1}\\
\vvbar_2\alpha_2&=\pibar_2\bv_2+\mu_1\rvec{\vbar_{10}&0&\vbar_{11}} +\mu_1\rvec{\vbar_{30}&0&\vbar_{31}},\eqnlabel{v-2nodes-q2}\\
\vvbar_{3}\alpha_{3}&=\pibar_3\bv_3
+\lambda\rvec{\vbar_{20}&0&\vbar_{22}}+\lambda\rvec{\vbar_{30}&0&\vbar_{32}}.
\end{align}
\end{subequations}
Since each $\vvbar_q$ has three components, there are twelve equations in \eqnref{AOI-SHS-v-2nodes}. However, because $x_1$ and $x_2$ are irrelevant in state $q=0$,  it follows from \eqnref{v-2nodes-q0} that $\vbar_{01}=\vbar_{02}=0$. Similarly, because $x_2$ is irrelevant in state $q=1$, and $x_1$ is irrelevant in state $q=2$, it follows from \eqnref{v-2nodes-q1} and \eqnref{v-2nodes-q2}  that $\vbar_{12}=\vbar_{21}=0$. Thus we are left with the eight equations
\begin{subequations}\eqnlabel{v-2nodes-8}
\begin{align}
\vbar_{00}\lambda &=\pibar_0+\mu_2\vbar_{22},\\
\vbar_{10}\mu_1&=\pibar_1+\lambda \vbar_{00}
+\mu_2\vbar_{32},\\
\vbar_{11}(\lambda+\mu_1)&=\pibar_1+\mu_2\vbar_{31},\\
\vbar_{20}(\lambda+\mu_2) &=\pibar_2+\mu_1\vbar_{10}+\mu_1\vbar_{30},\\
\vbar_{22}(\lambda+\mu_2) &=\pibar_2+\mu_1\vbar_{11}+\mu_1\vbar_{31},\\
\vbar_{30}(\mu_1+\mu_2)&=\pibar_{3}+\lambda \vbar_{20},\\
\vbar_{31}(\lambda+\mu_1+\mu_2)&=\pibar_{3},\\
\vbar_{32}(\mu_1+\mu_2)&=\pibar_{3}+\lambda \vbar_{22}.
\end{align}
\end{subequations}
With some algebra, it follows from  
\eqnref{v-2nodes-8} that
\begin{align}
\age(\lambda,\mu_1,\mu_2) &=\vbar_{00} +\vbar_{10} +\vbar_{20}+\vbar_{30}\nn
&=\paren{\frac{1}{\lambda}+\frac{1}{\mu_1}+\frac{1}{\mu_2}}
\sum_{q=0}^3\pibar_q\nn
&=\frac{1}{\lambda}+\frac{1}{\mu_1}+\frac{1}{\mu_2}.
\eqnlabel{age2nodes}
\end{align}
When we have 2 preemptive servers, \eqnref{age2nodes} verifies that the average age at the monitor is indeed insensitive to the ordering of the servers.
Moreover, in comparing \eqnref{age1node} and \eqnref{age2nodes}, one can see the simple pattern  that the $i$th node contributes $1/\mu_i$ to the age at the monitor. To verify this observation for an $n$-node network, however, the SHS method we employed for $n=2$ nodes tracks the state of each node in the network and thus requires the solution of $(n+1)2^n$ equations. In the next section, we show how to extend this simple pattern to $n$ nodes using an SHS that generates only $n+1$ equations.

\section{The $n$ Node Line Network: Fake Updates}\label{sec:fake}
A unusual feature of the LCFS-S server at each node is that tracking the  idle/busy state of a node is not actually essential because an arrival goes immediately into service whether or not an update is in service.  When node $i$ is busy, $x_i(t)$ encodes the age of the update in service. If that update completes service at time $t'$, it enters service at node $i+1$ and $x'_{i+1}=x_i$, whether or not it preempts any update that may have been in service at node $i+1$. To avoid tracking the idle/busy state at node $i$, when an update departs node $i$, we create a fake update at node $i$, with the same timestamp (and age $x_i(t)$) as the update that just departed. If a new update from node $i-1$ arrives at node $i$, it preempts the fake update and the fake update causes no delay to an arriving real update.  If the fake update does depart node $i$, it will go into service at node $i+1$, but it will have the same age as the update it will preempt at node $i+1$.

In \cite{YatesKaul2017arxiv}, fake updates are introduced in an SHS derivation of the average age of an LCFS-S queue in which the SHS has a single discrete state.  In this section, we show how to extend the fake updates approach to  prove the following theorem.
\begin{theorem}\thmlabel{line-age} With Poisson arrivals of rate $\lambda$ to  an $n$-node line network of LCFS-S servers with service rates $\seq{\mu}{1}{n}$, the average age at the monitor is
\begin{align*}
\age(\lambda,\mu_1,\ldots,\mu_n)=\frac{1}{\lambda}+\sum_{i=1}^n\frac{1}{\mu_i}.
\end{align*}  
\end{theorem}

\begin{IEEEproof}
Using fake updates, the Markov chain has the singleton state space $\Qcal=\set{0}$ with the trivial stationary distribution $\pibar_{0}=1$. As shown in Figure~\ref{fig:MC-n-nodes}, there is a $0\to0$ transition for each link $l$.  The transitions $l\in\set{0,1,\ldots,n}$ are shown in Table~\ref{tab:MC-n-nodes}. In the table, we omit the $q_l\to q'_l$ entry since it is $0\to0$ for each $l$. Note that 
\begin{itemize}
\item Transition $l=0$ marks the arrival of a fresh update at node $1$. In the continuous state $\xv$, we set $x'_1=0$ but all other  components of $\xv$ are unchanged.
\item In a transition $l\in\set{1,\ldots,n-1}$ an update departs node $l$ and arrives at node $l+1$.  At node $l+1$, $x'_{l+1}=x_l$.  At node $l$, $x'_l=x_l$ because a fake update begins service.
\item In transition $n$, an update departs node $n$ and is delivered to the monitor. The age at the monitor is reset to $x'_0=x_n$.
\end{itemize}
Each continuous-state age component $x_i(t)$, whether representing a real or fake update, ages at unit rate and thus 
$\bv_0=\rvec{1 & 1& \cdots & 1}$.
In the following, our notation will be simplified by defining $\mu_0=\lambda$. Applying \Thmref{AOI-SHS} to the transitions given in Table~\ref{tab:MC-n-nodes}, we obtain
\begin{align}
\vvbar_0\sum_{i=0}^n\mu_i
=\bv_0&+\mu_0\smrvec{\vbar_{00}&\vz&\vbar_{02}&\vbar_{03}&\cdots&\vbar_{0n}}\nn
&+\mu_1\smrvec{\vbar_{00}&\vbar_{01}&\vbar_{01}&\vbar_{03}&\cdots&\vbar_{0n}}\nn
&+\mu_2\smrvec{\vbar_{00}&\vbar_{01}&\vbar_{02}&\vbar_{02}&\cdots&\vbar_{0n}}\nn
&\qquad\vdots\nn
&+\mu_n\smrvec{\vbar_{0n}&\vbar_{01}&\vbar_{02}&\vbar_{03}&\cdots&\vbar_{0n}}.
\eqnlabel{v-n-nodes}
\end{align}
 It then follows from \eqnref{v-n-nodes} that
\begin{align}
\vbar_{01}\sum_{i=0}^n\mu_i=1+ \vbar_{01}\sum_{i=1}^n\mu_i,
\end{align}
implying $\vbar_{01}=1/\lambda$. In addition, for $k=2,3,\ldots,n$,
\begin{align}
\!\!\vbar_{0k}\sum_{i=0}^n\mu_i
&=1+\vbar_{0k}\sum_{i=0}^{k-2}\mu_i
+\mu_{k-1}\vbar_{0,k-1}+\vbar_{0k}\sum_{i=k}^{n}\mu_i.
\eqnlabel{v0k}
\end{align}
It follows from \eqnref{v0k} that
$\vbar_{0k} ={1}/{\mu_{k-1}}+\vbar_{0,k-1}$,
implying
\begin{align}\eqnlabel{v-k-nodes}
\vbar_{0k} &=\frac{1}{\lambda}+\sum_{i=1}^{k-1}\frac{1}{\mu_{i}},\qquad k=1,\ldots,n.
\end{align}
Finally, for component $\vbar_{00}$ of $\vv_0$, \eqnref{v-n-nodes} implies
\begin{align}
\vbar_{00}\sum_{i=0}^n\mu_i
&=1+\vbar_{00}\sum_{i=0}^{n-1}\mu_i
+\vbar_{0n}\mu_n.
\eqnlabel{v00}
\end{align}
This implies 
$\vbar_{00}={1}/{\mu_n} +\vbar_{0n}$. The claim then follows from \eqnref{v-k-nodes} and \Thmref{AOI-SHS}.
\end{IEEEproof}
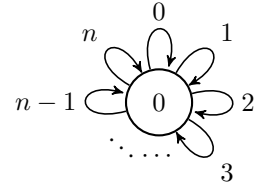
\begin{figure}
\centering
\begin{tikzpicture}[>=stealth', auto, semithick, node distance=2.75cm]
\tikzstyle{every state}=[fill=white,draw=black,thick,text=black,scale=1]
\node[state]    (0)                     {$0$};
\path
(0) 	edge[loop above,above]     node{$0$}  (0)
 edge[loop right,right]  node {$2$} (0)
(0) edge[loop left, left] node {$n-1$} (0);
\draw[->] (0) to [out=60,in=30,looseness=8] node {$1$} (0);
\draw[->] (0) to [out=330,in=300,looseness=8] node {$3$} (0);
\draw[->] (0) to [out=150,in=120,looseness=8] node {$n$} (0);
\draw ($(0)+(-0.5,-0.5)$) node {$\ddots$};
\draw ($(0)+(0,-0.72)$) node {$\cdots$};
\end{tikzpicture}
\caption{The SHS Markov chain for the line network with $n$ nodes.
The transition rates and transition/reset maps for links $l=0,\ldots,n$ are shown in Table~\ref{tab:MC-n-nodes}.}
\label{fig:MC-n-nodes}
\end{figure}
\begin{table}[t]
\begin{displaymath}
\setlength\arraycolsep{2pt}
\begin{array}{rccc}
l & \laml & \xv\Amat_l 
&\vvbar_{q_l}\Amat_l\\\hline
0  &\lambda 	& \rvec{x_0& \xz&x_2&x_3&\cdots& x_n} &\rvec{\vbar_{00}&\vz&\vbar_{02}&\vbar_{03}&\cdots&\vbar_{0n}}\\
1  &\mu_1 	& \rvec{x_0& x_1&x_1&x_3&\cdots& x_n} &\rvec{\vbar_{00}&\vbar_{01}&\vbar_{01}&\vbar_{03}&\cdots&\vbar_{0n}}\\
2  &\mu_2	& \rvec{x_0& x_1&x_2&x_2&\cdots& x_n} &\rvec{\vbar_{00}&\vbar_{01}&\vbar_{02}&\vbar_{02}&\cdots&\vbar_{0n}}\\
& \vdots & \vdots & \vdots\\
n  &\mu_n 	& \rvec{x_n& x_1&x_2&x_3&\cdots& x_n} &\rvec{\vbar_{0n}&\vbar_{01}&\vbar_{02}&\vbar_{03}&\cdots&\vbar_{0n}}\\
\end{array}
\end{displaymath}
\caption{Table of transitions for the Markov chain in Figure~\ref{fig:MC-n-nodes}.}\label{tab:MC-n-nodes}
\vspace{-5mm}
\end{table}

\section{Numerical Experiments}\label{sec:plots}
Because \Thmref{line-age} provides such a simple characterization of the average age in the preemptive line network, here we examine age sample paths in order to get a better sense of how age fluctuates across a sequence of preemptive servers. Our focus is  a line network with $n=3$ nodes.
In these experiments, let $\age_i(t)$ denote the instantaneous age at the output of node $i$ at time $t$. 

\begin{figure}[t]
\centering
\includegraphics{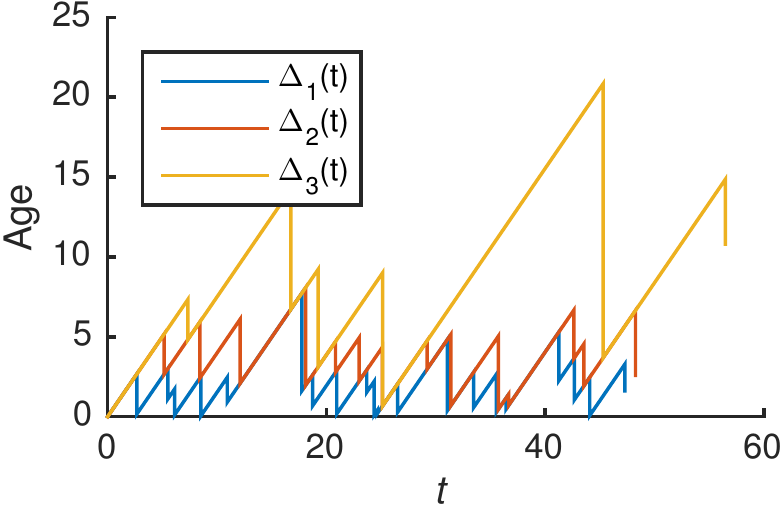}
\caption{Fifty arrivals to the three node line network with arrival rate $\lambda=1$ and service rates $\mu_1=1$, $\mu_2=0.5$ and $\mu_3=0.25$.}
\label{fig:ageplot}
\end{figure}

We start with Figure~\ref{fig:ageplot} which depicts representative sample paths  of $\age_i(t)$ for $i=1,2,3$. These sample paths are the result of  $50$ updates of a rate $1$ Poisson process arriving at node $1$ of a three-node network with service rates $\rvec{\mu_1 & \mu_2&\mu_3} = \rvec{1 & 0.5 & 0.25}$. These sample paths demonstrate that while the average age grows predictably, preemption thins the update arrival process at successive nodes. Because of this thinning, higher average age at successive nodes is a consequence of age growing in between less-frequent updates. We also see some evidence of survivor bias; occasionally some updates pass quickly through the network enabling the age at the monitor to be reset to a relatively low value. 

\begin{figure}[t]
\centering
\includegraphics{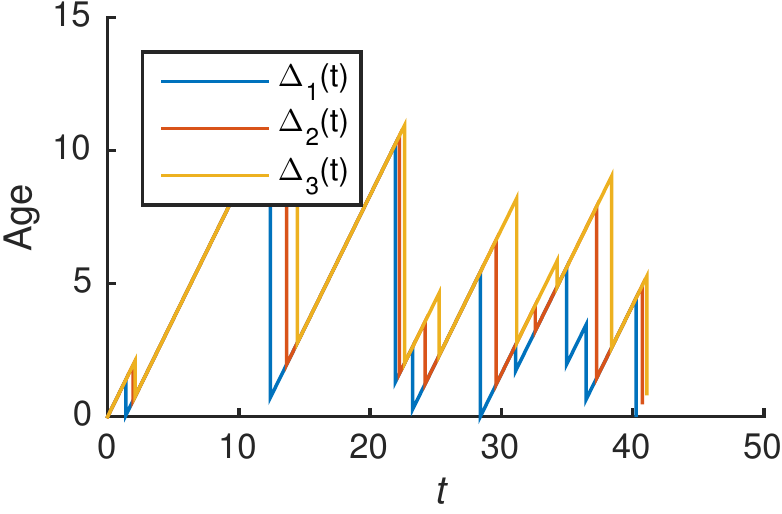}
\caption{Fifty arrivals to the three node line network with arrival rate $\lambda=1$ and service rates 
$\mu_1=0.25$, $\mu_2=0.5$ and $\mu_3=1$.}
\label{fig:ageplot2}
\end{figure}

In Figure~\ref{fig:ageplot2}, we examine the reverse  ordering of the servers; $\rvec{\mu_1 & \mu_2&\mu_3} = \rvec{0.25 & 0.5 & 1}$ so that downstream servers become progressively faster. In sample paths for this system,  most updates are dropped at the first node and the age processes at nodes $2$ and $3$ mimic the age at node $1$, but with some small additional lag. In this sample path, there is no evidence of survivor bias.  
However, it is apparent that in both Figures~\ref{fig:ageplot} and~\ref{fig:ageplot2} there is a general trend  that thinning of the updates induces larger age fluctuations at nodes further down the line. 

\begin{figure}[t]
\centering
\includegraphics{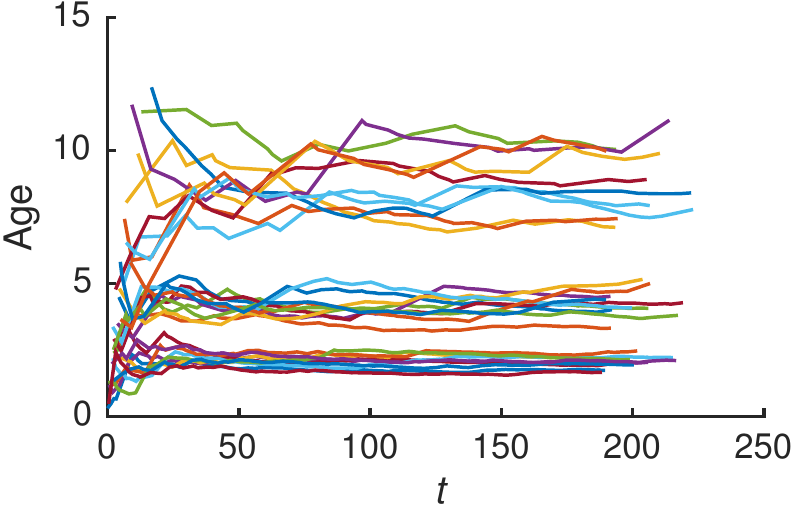}
\caption{Sample paths of the running sample average $\avgage_i(t)$, $i=1,2,3$, with arrival rate $\lambda=1$ and service rates $\mu_1=1$, $\mu_2=0.5$ and $\mu_3=0.25$. The bundles of sample paths converging to ages $2$, $4$, and $8$  support \Thmref{line-age}.}
\label{fig:avgagepaths}
\end{figure}

To see the convergence of the average age, we now examine  the running average age 
\begin{align}
\avgage_i(t)=\frac{1}{t}\int_0^t\age_i(\tau)\,d\tau
\end{align}
at each node $i$. In Figure~\ref{fig:avgagepaths}, we plot $10$ sample paths of $\avgage_i(t)$ for each node $i$. In each sample path, $200$ updates arrive at node $1$ as a rate $\lambda=1$ Poisson process. The nodes have service rates $\mu_1=1$, $\mu_2=0.5$ and $\mu_3=0.25$. For each sample path, we plot $\avgage_i(t)$ through the delivery time of the final received update at the monitor. From \Thmref{line-age}, we expect to see 
\begin{align}
\limty{t}\avgage_i(t) &=\frac{1}{\lambda}+\sum_{k=1}^i\frac{1}{\mu_k}=\begin{cases}
2, & i=1,\\
4 & i=2,\\
8 & i=3.
\end{cases}
\end{align}
In Figure~\ref{fig:avgagepaths}, we see that the bundles of running average sample paths clustered around ages $2$, $4$ and $8$ are  consistent with this conclusion. Moreover, the increasing variation and apparently slower convergence of the running average age in successive servers are also consistent 
the larger fluctuations in age at successive servers in the sample paths of Figures~\ref{fig:ageplot} and~\ref{fig:ageplot2}.

\section{Conclusions}\label{sec:conclusion}
In this work, we use the SHS approach to analyze age in an $n$-node line network of exponential servers. From the two node example in Section~\ref{sec:tandem}, it should be apparent  that SHS  can be employed to analyze a variety of other queues and simple networks described by finite continuous-time Markov chains.  Moreover, when each server in the network is a memoryless preemptive server, the method of fake updates can greatly simplify the age computation. We caution however that fake updates appear to be useful only when servers support preemption in service. For example, when preemptions occur in waiting, the discrete state must track whether a server is busy; i.e. it must distinguish between real updates and fake updates. In such cases, the system state will grow exponentially with the number of network nodes.  

\medskip

%

\bibliographystyle{unsrt}
\bibliography{paper}
\end{document}